\documentclass[a4paper,12pt,twoside,fleqn]{article}
\usepackage{epsfig,times,fancyheadings}
\usepackage{natbib,natbibspacing} 
\usepackage{tweaklist} 
\usepackage[normalem]{ulem} 

\normalsize 
\renewcommand{\enumhook}{%
  \setlength{\topsep}{0pt}%
  \setlength{\itemsep}{0pt}%
  \setlength{\parsep}{\parskip}} 
\makeatletter
\renewcommand{\section}{%
  \@startsection{section}{1}{0pt}{0pt}{1pt}{\centering}}
\renewcommand{\subsection}{%
  \@startsection{subsection}{2}{0pt}{0pt}{-1ex}{\uline}}
\renewcommand{\@seccntformat}[1]{{\csname the#1\endcsname}.~}
\makeatother
\renewcommand{\sectionmark}[1]{}
\renewcommand{\subsectionmark}[1]{}

\newcommand{\captionfonts}{\small}
\makeatletter  
\long\def\@makecaption#1#2{%
  \vskip\abovecaptionskip
  \sbox\@tempboxa{{\captionfonts #1: #2}}%
  \ifdim \wd\@tempboxa >\hsize
    {\captionfonts #1: #2\par}
  \else
    \hbox to\hsize{\hfil\box\@tempboxa\hfil}%
  \fi
  \vskip\belowcaptionskip}
\makeatother   

\begin{document}
\thispagestyle{plain}
\raisebox{0pt}[0pt][0pt]{\parbox[t]{0.9\linewidth}{\small
    \emph{Invited paper.}
    To be presented at:\\
    \textbf{The 19th Ris{\o} International Symposium on Materials
      Science:\\
      Modelling of Structure and Mechanics of Materials from
      Microstructure to Product}\\
    Ris{\o} National Laboratory, Roskilde, Denmark.
    }}
\vspace*{3cm}
\begin{center}
  SIMULATIONS OF MECHANICS AND STRUCTURE OF
  NANOMATERIALS --- FROM NANOSCALE TO COARSER SCALES\\[\baselineskip]
  J. Schi{\o}tz$^\ast$, T. Vegge, F. D. Di Tolla$^\dagger$ and
  K. W. Jacobsen\\[\baselineskip] 
  Center for Atomic-scale Materials Physics and Department of Physics,
  Technical University of Denmark, DK-2800 Lyngby, Denmark.\\
  $^\ast$\emph{Also at:} Materials Research Department, Ris{\o}
  National Laboratory, DK-4000 Roskilde, Denmark.\\
  $^\dagger$\emph{Present address:} SISSA, Via Beirut 2--4, I-34014
  Grignano (TS), Italy.
\end{center}

\section*{ABSTRACT}

We discuss how simulations of mechanical properties of materials
require descriptions at many different length scales --- from the
nanoscale where an atomic description is appropriate, through a
mesoscale where dislocation based descriptions may be useful, to
macroscopic length scales.  In some materials, such as nanocrystalline
metals, the range of length scales is compressed and a
polycrystalline material may be simulated at the atomic scale.  The
first part of the paper describes such simulations of nanocrystalline
copper.  We observe how the grain boundaries contribute actively to
the deformation.  At grain sizes below 10--15\,nm deformation in the
grain boundaries dominate over the traditional dislocation-based
deformation mechanisms.  This results in a reversal of the normal
grain size dependence of the yield stress: we observe that the
material becomes softer when the grain size is reduced.  The second
part of the paper gives an overview over simulation techniques
appropriate for problems too large to be treated in atomic-scale
simulations.  It also describes how different simulation techniques can
be combined to describe the interplay between phenomena at different
length scales through multiscale modelling.

\section{INTRODUCTION}

In the last decades the development of larger and faster computers has
progressed at a tremendous rate, doubling the capabilities of a
typical computer every 18 months.  This development has enabled the
field of ``computational materials physics'' to contribute
significantly to the understanding of materials and their properties.
In this paper we focus on the modelling of mechanical properties of
nanocrystalline metals, i.e. of metals with a grain size in the
nanometer range.  The main focus is on atomic-scale simulations, but
we also look at other simulation techniques for modelling materials at
coarser length-scales.

As matter is made of atoms, and as the quantum mechanical equations
governing the interactions of atoms (and the associated electrons) are
known, one could imagine that it is --- at least in principle ---
possible to solve these equations and predict the properties of matter
from first principles.  For simple properties of single-crystalline
defect-free metals this is indeed possible.
With quantum mechanical methods one is, however, only able to treat up
to a few hundred atoms, and even there one has to resort to
approximations when solving the fundamental equations.  When treating
complicated processes such as plastic deformation, this is
clearly inadequate.  One has to give up the ambition of
retaining a full quantum mechanical description of the atomic
interactions, and describe them by interatomic potentials.

Using parallel supercomputers, one can handle up to $10^8$ atoms for
times up to a nanosecond, when the interactions are described using
simple pair potentials (Abraham 1997; Abraham, Schneider, Land, Lifka,
Skovira, Gerner and Rosenkranz 1997; Bulatov, Abraham, Kubin,
Devincre, and Yip 1998)\nocite{Ab97,AbScLaLiSkGeRo97,BuAbKuDeYi98}, or
up to 35 million atoms using more realistic many-body potentials
(Zhou, Beazley, Lomdahl and Holian 1997)\nocite{ZhBeLoHo97}.  In most
cases one is limited to significantly smaller systems by the need to
run a large number of simulations (and often by economical factors as
well).  The length scale of typical processes in plastic deformation
is 1\,$\mu$m or more, and the time scale is typically seconds or
longer.  This corresponds to $10^{11}$ atoms in $10^9$ nanoseconds,
requiring a computational power of $10^{12}$ times what is possible
today.  Even if computers continue to improve at the current rate,
such computational power will not become available in a foreseeable
future.  One is therefore forced to abandon using an atomistic
approach to the whole problem, but will have to split up the problem
according to the different length scales involved, and reserve the
atomistic approach to the processes at the smallest length scales, see
e.g. Carlsson and Thomson (1998)\nocite{CaTh98}.

In this paper we will give an overview over how this gap between
different length scales can be bridged in simulations of plastic
behaviour.  In the first part we describe simulations of systems where
the characteristic length scales are so small that the entire system
can be studied atomistically.  The system we have chosen is
nanocrystalline metals.  In the second part we give an overview of
different simulation techniques at the micro and mesoscale, and how the
different scales can be combined.

\section{ATOMIC-SCALE SIMULATIONS OF NANOCRYSTALLINE METALS}

Nanocrystalline metals are metals with grain sizes on the nanometer
scale, typical grain diameters range from 5 to 50 nanometers.  These
materials are of technological interest, mainly because their strength
and hardness often are far above what is seen in coarse-grained
metals.  This is generally believed to be caused by the grain
boundaries acting as barriers to dislocation motion: as the grain size
is decreased, the number of grain boundaries increase and the
dislocation motion becomes harder, leading to a harder material.  This
grain size dependence of the hardness and the yield stress --- the
\emph{Hall-Petch effect} (Hall 1951; Petch 1953)\nocite{Ha51,Pe53} ---
is observed in coarse grained materials as well as in nanocrystalline
metals.  It is indeed possible to get a good estimate of the hardness
of nanocrystalline metals by extrapolating from the Hall-Petch
behaviour at larger grain sizes.  For a further discussion of the
mechanical properties of nanocrystalline metals, see e.g.\ the reviews
by Siegel and Fougere (1994)\nocite{SiFo94} or Morris and Morris
(1997)\nocite{MoMo97}.

Nanocrystalline metals are an attractive group of materials to model
for many reasons.  The materials are interesting from a technological
point of view, but also from a theoretical point of view since the
small grain size results in a ``cut off'' of the typical length scales
of the phenomena and structures that may appear during the deformation
process.  This simplifies the deformation process and may possibly
facilitate the development of theoretical models.  Eventually, it may
be possible to extend these models to the more complicated cases of
coarse grained materials.  For the smallest grain sizes (below
approximately 5--10\,nm) it becomes possible to model the deformation
process directly using atomic-scale computer simulations.  The
deformation processes may then be studied directly (Schi{\o}tz, Di
Tolla and Jacobsen 1998; Van Swygenhoven and Caro
1997a,b)\nocite{ScDiJa98,SwCa97,SwCa97b}.

\subsection{Generation of the ``samples''}

\begin{figure}[tbp]
  \begin{center}
    \leavevmode
    \epsfig{file=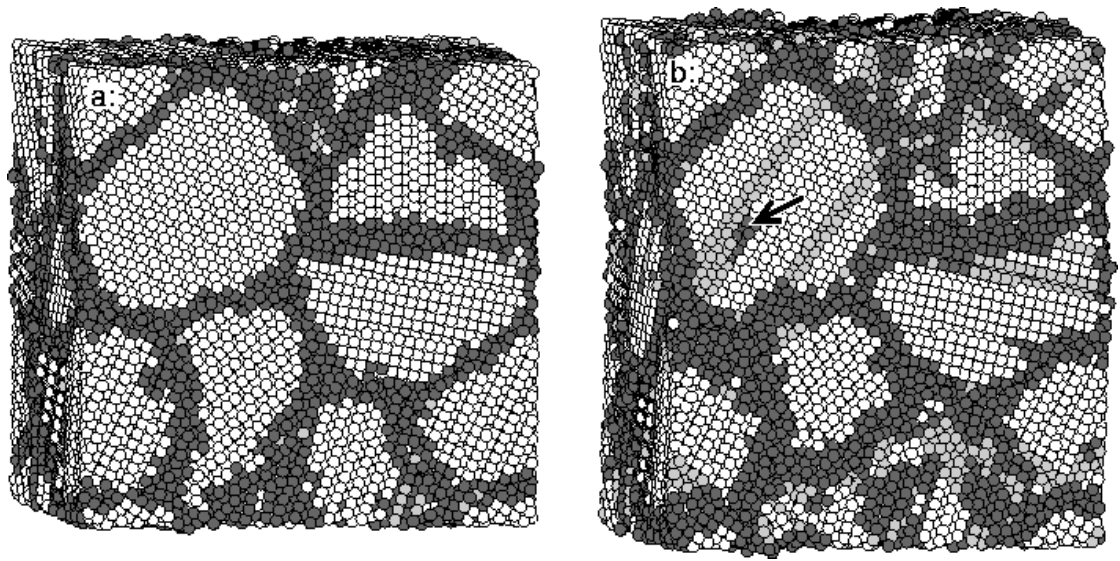, angle=-90, width=0.9\linewidth}
  \end{center}
  \caption[]{A simulated nanocrystalline copper sample before (a) and
    after (b) 10\% deformation.  The system contains approximately
    100,000 atoms arranged in 16 grains, giving an average grain
    diameter of 5.2\,nm.  Atoms are colour-coded according to the
    local crystalline order: white atoms are in a perfect f.c.c.\ 
    environment, light gray atoms are in a local h.c.p.\ environment,
    these atoms are at stacking faults.  Atoms in any other
    environment are coloured dark gray, these atoms are typically in
    grain boundaries or dislocation cores.  In the left side of the
    system, a partial dislocation is seen moving through a grain,
    leaving a stacking fault in its wake (arrow in (b)).  From
    Schi{\o}tz et al. (1998)\nocite{ScDiJa98}.}
  \label{fig:beforeafter}
\end{figure}

We attempt to generate ``samples'' with a structure reasonably similar
to the structures observed experimentally: essentially equiaxed
dislocation free grains separated by narrow grain boundaries.  The
grains are produced using a Voronoi construction.  A set of
grain centres are chosen randomly, and the part of space closer to a
given centre than to any other centre is filled with atoms in a
randomly rotated f.c.c.\ lattice.  Periodic boundary conditions are
imposed.  This procedure generates samples without texture, and with
random grain boundaries.  In the grain boundaries thus generated, it
is possible that two atoms from two different grains get too close to
each other, in such cases one of the atoms is removed to prevent
unphysically large energies and forces as the simulation is started.
To obtain more relaxed grain boundaries the system is annealed for
10\,000 timesteps (50\,ps) at 300\,K, followed by an energy
minimisation.  This procedure is important to allow unfavourable local
atomic configurations to relax.  A sample generated in this way is
shown in Fig.~\ref{fig:beforeafter}.

To investigate whether the parameters of the annealing procedure are
critical, we have annealed the same sample for 50 and 100\,ps at
300\,K, and for 50\,ps at 600\,K.  We have compared the mechanical
properties of these samples with those of an identical sample without
annealing, we find that the annealing is important (the unannealed
sample was softer), but the parameters of the annealing are
not important within the parameter space investigated.

A similar generation procedure has been reported by Chen
(1995)\nocite{Ch95}, by D'Agostino and Van Swy\-gen\-hoven
(1996)\nocite{AgSw96}, and by Van Swygenhoven and Caro
(1997a,b)\nocite{SwCa97,SwCa97b}.  A different approach was proposed
by Phillpot, Wolf and Gleiter (1995a,b)\nocite{PhWoGl95,PhWoGl95b}: a
nanocrystalline metal is generated by a computer simulation where a
liquid is solidified in the presence of crystal nuclei, i.e.\ small
spheres of atoms held fixed in crystalline positions.  The system was
then quenched, and the liquid crystallised around the seeds, thus
creating a nanocrystalline metal.  In the reported simulations, the
positions and orientations of the seeds were deterministically chosen
to produce eight grains of equal size and with known grain boundaries,
but the method can naturally be modified to allow randomly placed and
oriented seeds.  The main drawback of this procedure is the large
number of defects (mainly stacking faults) introduced in the grains by
the rapid solidification.  The stacking faults are clearly seen in the
resulting nanocrystalline metal (Fig.~7 of Phillpot et al.\ 1995a).
Some of us have earlier performed simulations of the solidification of
a large cluster (unpublished).  These simulations have shown that a
large number of stacking faults appear even if the cooling is done as
as slowly as possible in atomistic simulations.

\subsection{The simulation technique.}

We model the interactions between the atoms using a many-body
potential known as the Effective Medium Theory (EMT) (Jacobsen,
N{\o}rskov and Puska 1987; Jacobsen, Stoltze and N{\o}rskov
1996)\nocite{JaNoPu87,JaStNo96}.  It is very important that the
interactions are modelled using a realistic many-body potential such
as EMT, the Embedded Atom Method (Daw and Baskes 1984)\nocite{DaBa84}
or the Finnis-Sinclair model (Finnis and Sinclair 1984,
1986)\nocite{FiSi84,FiSi86}.  Pair potentials such as Lennard-Jones
are still seen used for simulating metals due to the lower
computational burden.\footnote{Typically, many-body potentials require
approximately ten times as much computer power as Lennard-Jones
potentials.}  Although they give a good description of noble gas
solid, they are not adequate for modelling the bonding in metallic
systems.  One symptom of this is seen in the elastic constants, all
pair potentials result in materials that satisfies the Cauchy
relations between the elastic constants ($C_{12} = C_{44}$ for cubic
crystals), a relation that is far from true in most metals.

In the simulations, the samples are deformed by slowly increasing the
system size along the $z$ axis while minimising the energy with
respect to all atomic coordinates and with respect to the box
dimensions in the $x$ and $y$ directions.  The minimisation is done as
a modified molecular dynamics simulation.  After each timestep the dot
product between the momentum and the force is calculated for each
atom.  Any atom where the dot product is negative gets its momentum
zeroed, as it is moving in a direction where the potential energy is
increasing.  This \emph{MDmin} algorithm (Stoltze 1997)\nocite{St97}
is very efficient for this type of minimisation.  Before each
timestep, the system is stretched a little along the $z$ direction.
The two lateral dimensions are optimised by a Monte Carlo procedure:
every 20 timesteps a change in the dimensions is proposed, if the
change results in a lower energy it is accepted, otherwise it is
discarded.  A few simulations were performed using the conjugate
gradient algorithm for energy minimisation (Press, Flannery, Teukolsky
and Vetterling 1988)\nocite{PrFlTeVe88}.  The two algorithms were
approximately equally efficient.

\subsection{Analysis.}

\begin{figure}[tbp]
  \begin{center}
    \begin{minipage}[t]{0.58\linewidth}
      \epsfig{file=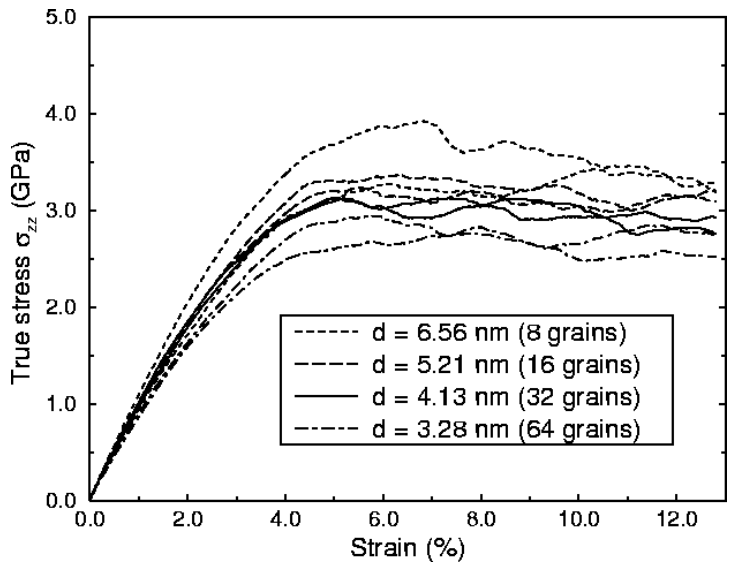, angle=-90, width=\linewidth}
    \end{minipage}\hfill
    \begin{minipage}[t]{0.38\linewidth}
      \caption{Stress-strain curves for eight simulations with four
        different grain sizes $d$.  A total of 28 simulations were
        performed, but for clarity only eight are shown here.  The
        fluctuations in the individual curves and the differences
        between curves at the same grain size, are due to the low
        number of grains in each system.\label{fig:rawss}}
    \end{minipage}
  \end{center}
\end{figure}

During the simulation, the local stresses were calculated and stored
for further analysis, and the global values of the stress tensor were
stored to generate stress-strain curves.  A set of stress-strain
curves is shown in Fig.~\ref{fig:rawss}.

To facilitate the analysis of the simulations, the local order in the
sample was analysed.  This was done using a method called Common
Neighbour Analysis (J\'onsson and Andersen 1988; Clark and J\'onsson
1993)\nocite{JoAn88,ClJo93}.  By investigating the bonds between the
neighbouring atoms to the atom under investigation, the local crystal
structure is determined.  Atoms are then classified into three
classes.  Atoms in a local f.c.c.\ order are considered ordinary bulk
atoms; atoms in local h.c.p.\ order are labelled as belonging to a
stacking fault or the like; and atoms in all other local orders are
considered part of the grain boundaries or of dislocation cores.  This
classification can be seen in Fig.~\ref{fig:beforeafter}.

\subsection{Results.}

\begin{figure}[tbp]
  \begin{center}
    \leavevmode
    \epsfig{file=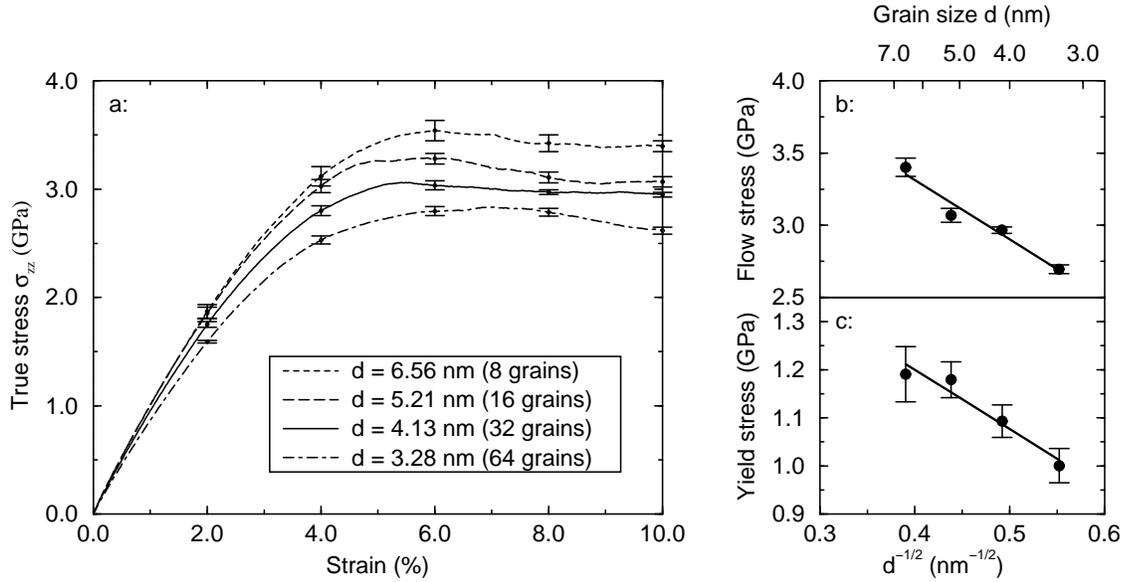, angle=-90, width=0.9\linewidth}
  \end{center}
  \caption{(a)  The average stress versus strain for
    each grain size.  Each curve is the average over seven
    simulations.  The curves show the response of the material to
    mechanical deformation.  In the linear part of the curve (strains
    less than 1--2\%) the deformation is mainly elastic.  As the
    deformation is increased irreversible plastic deformation becomes
    important.  For large deformations plastic processes relieve the
    stress, and the curves level off.  We see a clear dependence on
    the grain size $d$, it is summarised to the right.  (b) and (c)
    The maximal flow stress and the yield stress as a function of
    grain size.  The yield stress decreases with decreasing grain
    size, resulting in a \emph{reverse} Hall-Petch effect.  The
    maximal flow stress is the stress at the flat part of the
    stress-strain curves.  The yield stress is defined as the stress
    where the strain departs 0.2\% from linearity.  Adapted from
    Schi{\o}tz et al.\ (1998).}
  \label{fig:hallpetch}
\end{figure}

Fig.~\ref{fig:hallpetch}a shows the average stress-strain curves at
each grain size.  A clear grain size dependence is seen in the maximal
flow stress.  Fig.~\ref{fig:hallpetch}b and c summarise the results,
showing the maximal flow stress and the 0.2\% offset yield stress in a
standard Hall-Petch plot, i.e.\ as a function of one over the square
root of the grain size.  A reverse Hall-Petch effect is clearly seen,
it will be discussed in section \ref{sec:revhp}.

In Fig.~\ref{fig:beforeafter}b stacking faults are present in several
grains.  They are created as partial dislocations (Shockley partials)
move through the grains.  They would be removed if a second partial
dislocation followed in the same plane.  This is, however, rarely seen,
possibly because the grain size is comparable to the splitting width
of the dislocation.  Simulations of dislocation emission from a notch
in a surface (Schi{\o}tz, Jacobsen and Nielsen 1995)\nocite{ScJaNi95}
have shown that the leading partial will move several splitting
widths away from the notch, before the trailing partial is emitted.
Possibly a similar effect is seen here.

\begin{figure}[tbp]
  \begin{center}
    \leavevmode
    \epsfig{file=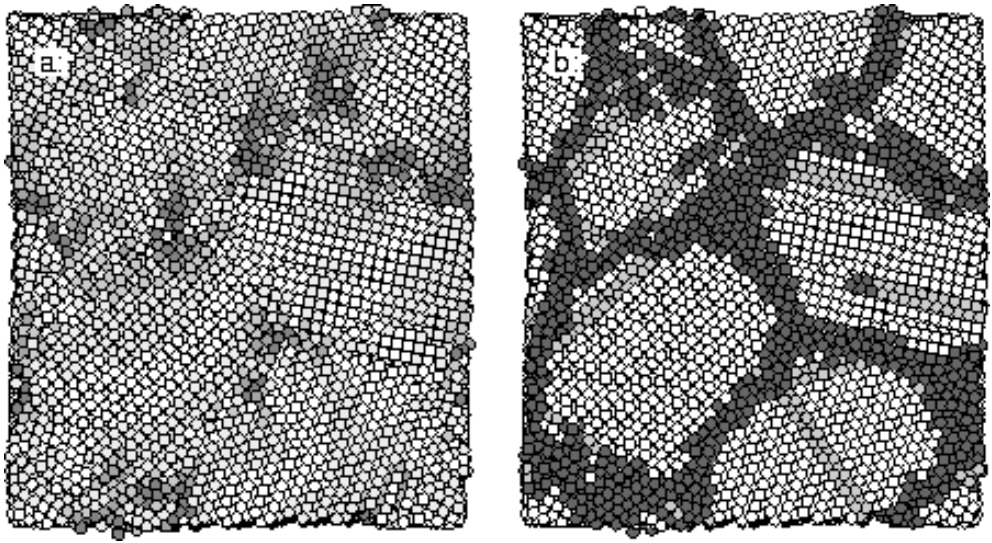, angle=-90, width=0.8\linewidth}
  \end{center}
  \caption[]{(a) The relative motion of the atoms in the $z$ direction
    (up in the plane of the paper) during the preceding 0.4\%
    deformation.  Darker atoms move more than light atoms.  We see
    many small, independent slip events in the grain boundaries, this
    is the main deformation mode. (b) The position of the grain
    boundaries and stacking faults at this point in the simulation.
    Adapted from Schi{\o}tz et al.\ (1998).}
  \label{fig:slip}
\end{figure}

We have measured the amount of dislocation activity in the systems.
The total dislocation activity is able to explain at most 2--3\%
plastic deformation, i.e. less than half the observed plastic
deformation.  To identify the nature of the remaining deformation, we
analysed the relative motion of the individual atoms.
Fig.~\ref{fig:slip}a shows the magnitude of the motion of the atoms
obtained by subtracting the coordinates at two different strains
(differing by 0.4\%).  The collective motion due to the imposed strain
rate has been subtracted out.  Darker atoms move most.  By comparing
with Fig.~\ref{fig:slip}b showing the positions of the grain
boundaries, it is seen that the motion occurs mainly in the grain
boundaries.  An indication of the activity in the grain boundaries can
also be seen in Fig.~\ref{fig:beforeafter}: the grain boundaries have
become a little thicker during deformation.  

\section{THE REVERSE HALL-PETCH EFFECT}
\label{sec:revhp}

As is seen in Fig.~\ref{fig:hallpetch}, a reverse Hall-Petch effect is
observed in the simulations, i.e.\ the material gets softer as the
grain size is decreased.  This is opposite of the normally seen
behaviour, but has occasionally been observed experimentally in
nanocrystalline materials with sufficiently small grain sizes
(Chokshi, Rosen, Karch, Gleiter 1989)\nocite{ChRoKaGl89}.  There are
also observations of ``kinked'' Hall-Petch graphs, i.e.\ cases where
the slope is reduced (but still positive) below a certain grain size.
Many explanations have been proposed, of which some are summarised
below.  The explanations are not necessarily mutually exclusive, nor
is it unlikely that different explanations may apply depending on the
way the material was produced.  It seems reasonable to assume that the
deformation mechanism is different in samples produced by e.g.\ inert
gas condensation, producing essentially dislocation-free grains, and
by severe plastic deformation, where the small grains are produced by
breaking up larger grains through intense dislocation activity.  The
most commonly proposed explanations are increased diffusional creep,
suppression of dislocation pileups, different grain boundary
structures, poor sample quality (porosity and other flaws), and
deformation in the grain boundaries.

\subsection{Enhanced Coble creep.}

Chokshi et al.\ (1989) propose that the reverse Hall-Petch effect is
caused by enhanced Coble creep, i.e.\ creep due to diffusion in the
grain boundaries.  Coble creep scales with the grain size ($d$) as
$d^{-3}$, and estimates of the creep rate of nanocrystalline metals
indicate that this could be the explanation of the reverse Hall-Petch
effect.  Direct measurements of the creep rate have, however, ruled
out this explanation (Nieman, Weertman and Siegel 1990; Nieh and
Wadsworth 1991)\nocite{NiWeSi90,NiWa91}.

\subsection{Suppression of dislocation pileups.}

The Hall-Petch effect is normally explained by appealing to
dislocation pileups near the grain boundaries.  Once the grain size
drops below the equilibrium distance between dislocations in a pileup,
pileups are no longer possible, and the Hall-Petch relation should
cease to be valid (Nieh and Wadsworth 1991; Pande, Masumura and
Armstrong 1993)\nocite{NiWa91,PaMaAr93}.  It is, however, not clear
how the yield stress should depend on the grain size below that point.
The critical grain size is estimated to be 20\,nm for copper (Nieh and
Wadsworth 1991).

\subsection{Different grain boundary structures.}

Is is not unreasonable to assume that the grain boundary structure
might be different when the grain size is very small (Zhu, Birringer,
Herr and Gleiter, 1987)\nocite{ZhBiHeGl87}.  It has been proposed that
grain boundaries in nanocrystalline metals may be more ``transparent''
to dislocations, and thus allow dislocations to run through several
grains (Valiev, Chmelik, Bordeaux, Kapelski and Baudelet 1992; Lian,
Baudelet and Nazarov 1993; Lu and Sui
1993)\nocite{VaChBoKaBa92,LiBaNa93,LuSu93}.  This was proposed in
connection with measurements of a possible breakdown of the Hall-Petch
relation in metals with sub-micrometer grain sizes, produced by severe
plastic deformation (Valiev et al. 1992).  Recent high-resolution
electron microscopy studies show, that the grain boundaries in metals
produced in this way have a complex structure, with a large number of
dislocations very close to the grain boundary (Horita, Smith,
Furukawa, Nemoto, Valiev and Langdon 1996)\nocite{HoSmFuNeVaLa96}.
This should make the grain boundaries \emph{less} transparent to
dislocations, but a change in slope in the Hall-Petch relation is seen
at grain sizes below 100\,nm (Furukawa, Horita, Nemoto, Valiev and
Langdon 1996)\nocite{FuHoNeVaLa96}.  They explain the change as
enhanced plasticity due to these extra dislocations near the grain
boundaries.

\subsection{Porosity and flaws.}

Many of the observations of a reverse Hall-Petch effect are from
samples generated using inert gas condensation, where a large number
of nanometer-sized clusters are compacted to produce the sample.  If
the compaction is not complete, small voids will be present between
the grains.  The presence of these voids was not initially recognised,
the lower density being ascribed to special low-density grain
boundaries.  If the nanocrystalline metal contains a significant
volume fraction of porosity, this will obviously reduce the hardness
significantly.  Surface defects alone have been shown to be able to
reduce the strength of nanocrystalline metals by a factor of five
(Nieman, Weertman and Siegel 1991; Weertman
1993)\nocite{NiWeSi91,We93}.

Many of the early measurements of a reverse Hall-Petch
effect are likely to have been caused by unrecognised porosity in the
samples.  Improved techniques (Sanders, Fougere, Thompson, Eastman and
Weertman 1997)\nocite{SaFoThEaWe97} have allowed production of
nanocrystalline samples with densities above 98\%, these have shown no
reverse Hall-Petch effect in copper at grain sized down to approximately
10--15\,nm (Sanders, Youngdahl and Weertman 1997)\nocite{SaYoWe97}.

\subsection{Deformation in the grain boundaries.}

As the grain size is reduced, the volume fraction of the grain
boundaries increase, and it is reasonable to assume that at some point
they will begin to play a role in the deformation process.  Li, Sun
and Wang (1994)\nocite{LiSuWa94} propose a deformation mechanism based
on motion of grain boundary dislocations.

\subsection{The role of computer simulations.}

\begin{figure}[tbp]
  \begin{center}
    \begin{minipage}[t]{0.58\linewidth}
      \epsfig{file=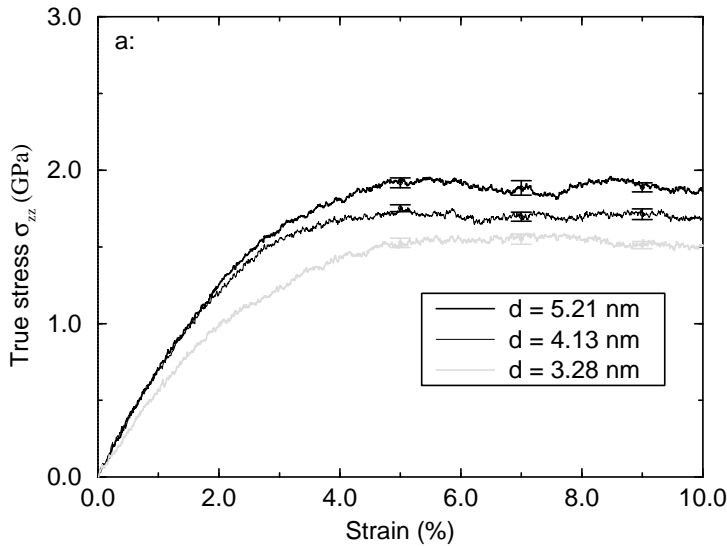, angle=-90, width=\linewidth}
    \end{minipage}\hfill
    \begin{minipage}[t]{0.38\linewidth}
      \caption{Stress stress curves for simulations at room
        temperature.  Each curve shows the average over four
        simulations.  The simulations were done using constant-energy
        molecular dynamics at a strain rate of $5 \cdot 10^8
        \mbox{s}^{-1}$.  $d$ is the grain size.}
      \label{fig:roomtemp}
    \end{minipage}
  \end{center}
\end{figure}

\begin{sloppypar}
  Computer simulations give a possibility to distinguish between this
  large number of different explanations.  The computer simulations
  presented here clearly indicate that a new deformation mechanism
  becomes active in the grain boundaries (Schi{\o}tz et al. 1998).  It
  does not appear to be a grain boundary dislocation based motion, but
  rather a large number of small events, where only a few atoms (or a
  few tens of atoms) move simultaneously (Fig.~\ref{fig:slip}). This
  produces a large number of small, apparently uncorrelated slipping
  events in the grain boundaries, leading to grain boundary sliding.
  The processes also lead to the grain boundaries becoming thicker
  (Fig.~\ref{fig:beforeafter}).  This does, however, seem to be a
  result of doing the simulations at zero temperature.  Simulations at
  room temperature do not result in a change in the grain boundary
  thickness, otherwise the results are very similar apart from a
  reduction of the yield and flow stresses due to increased grain
  boundary sliding (Fig.~\ref{fig:roomtemp}).
\end{sloppypar}

Simulations of nanocrystalline nickel also show that the main
deformation mechanism is grain boundary sliding and grain boundary
motion, where one grain grows at the expense of another (Van
Swygenhoven and Caro, 1997a,b)\nocite{SwCa97,SwCa97b}.  Those
simulations are stress controlled, i.e.\ the stress is controlled, and
the strain is measured, contrary to the simulations reported here,
which are strain controlled.  The strain rate was measured as a
function of grain size, giving a higher strain rate at lower grain
boundaries, consistent with a viscoelastic behaviour of the grain
boundaries (Van Swygenhoven and Caro, 1997b).

Simulations of the sintering of nanocrystalline Cu particles to form a
fully dense nanocrystalline sample also indicate that the main
deformation is in the grain boundaries (Zhu and Averback,
1996)\nocite{ZhAv96}.

\begin{figure}[tbp]
  \begin{center}
    \leavevmode
    \epsfig{file=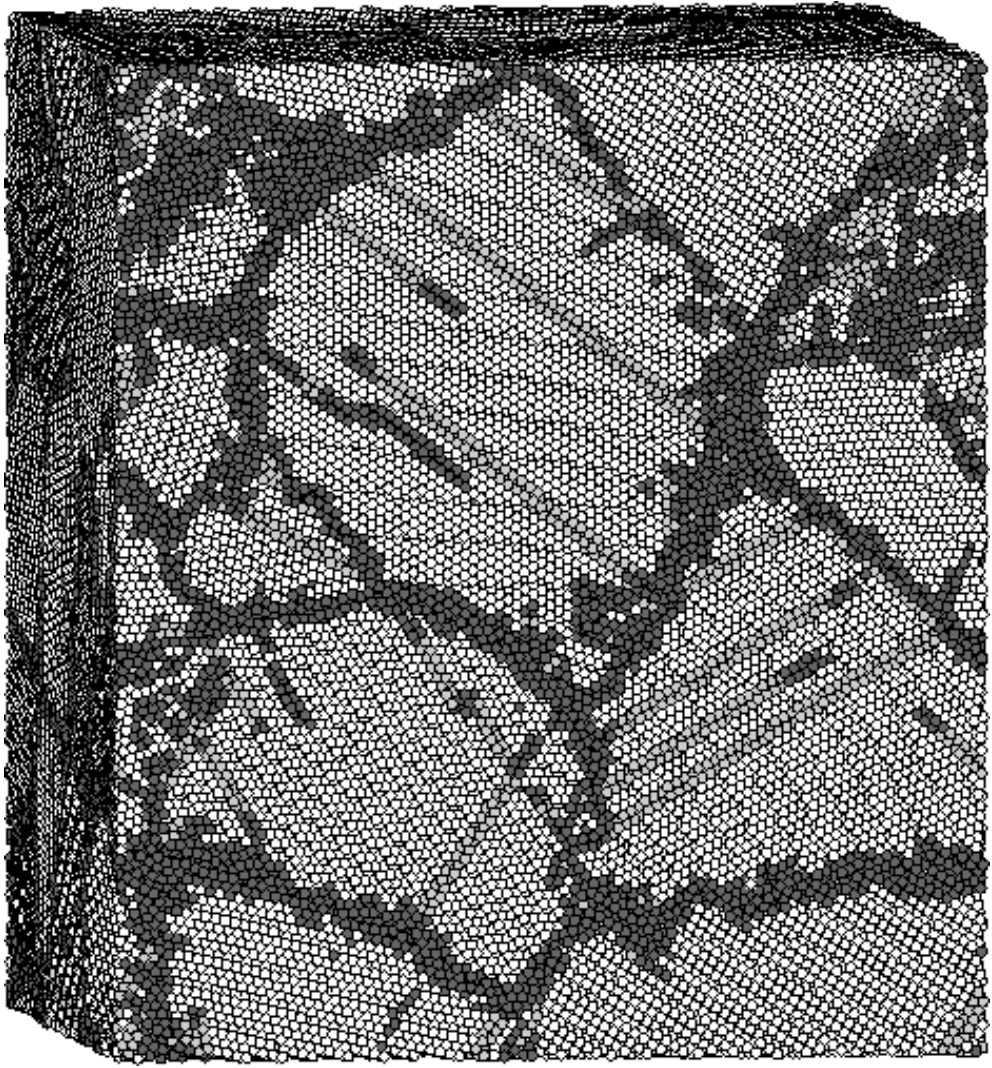, clip=, width=0.8\linewidth}
  \end{center}
  \caption{A simulated nanocrystalline copper sample after 10\%
    deformation.  The grain size is 13.2\,nm, and the system contains
    1,009,474 atoms.}
  \label{fig:bigsystem}
\end{figure}

The observations of a reverse Hall-Petch effect in the simulations
presented here are not in conflict with the experiments reported by
Sanders et al.\ (1997)\nocite{SaYoWe97}, since the experiments were
done on samples with grain sizes above 10--15\,nm.  In order to
investigate if the transition between the normal and the reverse
Hall-Petch effect could be observed, a version of the simulation
program was written for parallel computers.  This allowed us to
increase the number of atoms by a factor of ten, thus more than
doubling the grain size.  Such a simulation is shown in
Fig.~\ref{fig:bigsystem}.  Preliminary results indicate that the
reverse Hall-Petch effect persists at these grain sizes.

\section{BEYOND THE ATOMIC SCALE}

As larger systems are being considered, atomic scale simulations
quickly become impossible: doubling the characteristic length scale
typically means increasing the volume, and thus the number of atoms,
by a factor eight.  At present, simulations with up to 100 million
atoms are possible using parallel supercomputers (Abraham 1997;
Abraham \emph{et al.}\ 1997; Zhou \emph{et al.}\
1997)\nocite{Ab97,AbScLaLiSkGeRo97,ZhBeLoHo97}, corresponding to a
system of less than 100\,nm cubed.  If a larger system is to be
studied, simulations which consider individual atoms cannot in
practise be carried out.  In this section we give a short summary over
a few methods that go beyond an atomic description to describe the
material at a coarser length scale.  See also the review by Carlsson
and Thomson (1998)\nocite{CaTh98}, where simulation techniques at
different length scales are discussed in the context of how to
calculate fracture toughness.

\subsection{Simulations based on individual dislocations.}

The idea behind ``dislocation dynamics'' is to use individual
dislocations as ``fundamental particles'' in the simulations, i.e.\ to
describe the material as consisting of a collections of dislocations
that interact with each other through their elastic fields.  This is a
sensible idea when studying properties dominated by the dislocation
dynamics, but cannot be expected to work without modifications in
cases where other defect types (e.g.\ cracks and grain boundaries)
play a significant role.

In principle the methods developed for atomistic simulations can be
used, provided one replaces the interatomic interactions with the
formula for interactions between dislocations.  There are, however, a
number of complications:
\begin{enumerate}
\item Dislocations are lines, not points, i.e.\ the position of a
  dislocation cannot be described by a single set of coordinates.
\item Both the force acting between dislocations, and the equation of
  motion for a single dislocation are highly anisotropic.
\item Interactions between dislocations are long ranged, the force
  decays as $1/r$, whereas the forces between atoms in a metal decay
  exponentially due to screening effects.  It is thus not possible to
  introduce a cutoff distance beyond which dislocations do not
  interact.  Attempts to introduce such cutoffs have been shown to
  cause the emergence of spurious structures with a characteristic
  length scale equal to the chosen cutoff (Gulluoglu, Srolovitz, LeSar
  and Lomdahl 1989; Holt 1970)\nocite{GuSrLeLo89,Ho70}.
\item Although the long-range elastic forces between dislocations are
  well known, the parameters describing the short-range interactions
  (dislocation crossing, cross slip, annihilation etc.) are presently
  not well known, and will have to be extracted from experiments,
  come from theoretical arguments or come from atomic-scale simulations.
\end{enumerate}

Barts and Carlsson (1995)\nocite{BaCa95} have developed a method for
simulating the interactions of a large number of dislocations in
two-dimensional systems.  In two dimensions the dislocations are
points, eliminating the first problem mentioned above, and simplifying
the interactions significantly.  The drawback is of course that the
simulations are limited to problems that are essentially
two-dimensional, i.e. where the dislocations are parallel and
straight.  One such problem, which has been successfully modelled, is
polygonisation in single glide (Barts and Carlsson
1997)\nocite{BaCa97}.  They solve the problem of the long range
interactions between dislocations by using a Fourier transform of the
long range part of the interactions, resulting in a method where the
required computer time scales linearly with the number of dislocations.

Similar methods can be developed in three dimensions, as demonstrated
by Kubin and coworkers (Kubin, Canova, Condat, Devincre, Pontikis and
Br\'echet 1992; Devincre and Kubin
1997)\nocite{KuCaCoDePoBr92,DeKu97}.  They solve the problem of
representing the position of the dislocation strings by modelling the
dislocations as consisting of a chain of straight segments of pure
screw or edge dislocations, where each segment can be oriented along a
finite number of directions.  The interactions between the
dislocations are then described by the interactions of the individual
segments.  The method has been used to study the influence of cross
slip on work hardening (Devincre and Kubin 1994)\nocite{DeKu94} and on
pattern formation during cyclic deformation (Kratochv{\'\i}l,
Saxlov\`a, Devincre and Kubin 1997)\nocite{KrSaDeKu97}.

\subsection{Simulations based on dislocation densities.}

When the structures under consideration become larger, even
dislocation based simulations become unwieldy.  It is tempting to
attempt to make a model based on a continuum description of the
dislocation density, i.e.\ by replacing the positions of individual
dislocations by a continuous dislocation density.  A simple scalar
density is clearly insufficient to describe the dislocations; as a
dislocation is described by two vectors (the sense and the Burgers
vector), the density becomes a tensor (Nabarro 1987)\nocite{Na87}.

As shown by Barts and Carlsson (1997)\nocite{BaCa97} a continuum
description based on the dislocation density tensor alone is
insufficient to describe a relatively simple patterning process such as
polygonisation.  This is because the stress field from a dislocation
wall is caused by the discrete dislocation positions, and vanish in a
continuum model.  They propose that a continuum model should contain
additional variables describing the local environment of the
dislocations.  It is at present unclear if such a description can be
made.

For an overview of the work that has been done on continuum modelling
of dislocations, see the review by Selitser and Morris
(1994)\nocite{SeMo94}.

\subsection{Simulations based on grains.}

For some types of simulations, an even coarser length scale can be
used: the scale set by the size of the gains.  In typical metals this
will be of the order of 100\,$\mu$m.  At this length scale one may
consider the grains as the fundamental unit of the simulation.  This
approach has successfully been used e.g.\ to study recrystallization
and the evolution of texture during growth (Juul Jensen 1997a,b and
references therein)\nocite{Ju97a,Ju97b}.

\subsection{Finite-element methods.}

For engineering purposes, the relevant length scale will be
set by the physical dimensions of the sample.  Typically simulations
ignore the grain structure of the material, which is treated as 
homogeneous (but possibly anisotropic).  A continuum description based
on the stress and strain fields is used, and the field equations are
typically solved using finite element methods.

\section{BRIDGING THE LENGTH SCALES}

To gain a fuller understanding of material properties, modelling at
several length scales should be combined (often referred to as
``multiscale modelling'').  Typically, experiments or simulations at
one length scale suggests interesting phenomena to study at a shorter
length scale, the result of which is used as parameters in models at a
longer length scale.  In the following we will give some examples,
followed by some examples of ``hybrid'' simulation techniques, where
several simulation paradigms are combined in a single simulation.

\subsection{Interactions between simulations at different length
  scales.}

Dislocation level simulations require the knowledge of details of the
short-range interactions between dislocations.  Many of these can come
from atomic-scale simulations.  This can be done by setting up a very
large system, and observe the dislocation processes in this system.
Bulatov \emph{et al.}\ (1998)\nocite{BuAbKuDeYi98} have studied the
behaviour of the cloud of dislocations emitted from a crack in an
f.c.c. metal, and observed the creation and subsequent destruction of
a Lomer-Cotrell lock.  They propose to extract parameters for
dislocation level simulations from such atomistic simulations.
Schi{\o}tz, Jacobsen, and Nielsen (1995)\nocite{ScJaNi95} have
observed a new dislocation multiplication mechanism that may be active
under extremely large strain rates.

If the main goal of the simulation is to provide knowledge of a
specific dislocation process, a more specialised simulation setup may
be advantageous.  Zhou, Preston, Lomdahl and Beazley
(1998)\nocite{ZhPrLoBe98} have simulated the formation of a jog
through the intersection of two extended dislocations.  The
simulations show the details of the interaction and provides an
estimate of the involved critical stresses.

Recently an advanced simulation technique, the ``Nudged Elastic Band
(NEB) Method'' (Mills, J\'onsson and Schenter 1995)\nocite{MiJoSc95}
was used to study cross slip of screw dislocations at the atomic scale
(Rasmussen, Jacobsen, Leffers, Pedersen, Srinivasan and J\'onsson
1997; Pedersen, Carstensen and Rasmussen
1998)\nocite{RaJaLePeSrJo97,PeCaRa98x}.  The NEB method identifies the
entire transition path and can thus be used to find the energy of the
transition state.  This makes the method very suitable for studying
elementary dislocation reactions, and can provide good input for
dislocation level simulations.  The group is presently using the
technique to study dislocation annihilation (Pedersen et al.
1998)\nocite{PeCaRa98x}.

\subsection{Hybrid simulation techniques.}

A number of simulation techniques have been developed, where several
simulation techniques are combined in a ``hybrid'' simulation.
Typically, the idea is to describe ``interesting'' parts of the
simulation atomistically,  while other parts of the system are
described by a model based on linear elasticity theory.  The rationale
behind this is that in most large-scale simulations, the majority of
the atoms do not participate actively in the processes being studied,
their main role is to propagate the elastic fields.  In these regions
of the simulation the strains are typically small, and linear
elasticity is a good description.  At the same time an atomistic
description is required in dislocation core and other parts of the
system, where non-linearities play a role.

The simplest way of doing this is by dividing the simulations into two
zones.  In the inner zone an atomistic description is used.  The inner
zone is surrounded by a much larger outer zone, described by
elasticity theory typically using a continuous finite element method
(Kohlhoff, Gumbsch and Fishmeister 1991; Gumbsch
1995)\nocite{KoGuFi91,Gu95}.  In this way the long-range elastic field
is described in a computationally inexpensive way, and the atoms in
the atomistic region see the correct response from the surrounding
material.  Great care must be taken in matching the two regions.

A related method consists of using a lattice Green's function to
describe the linear region (Thomson, Zhou, Carlsson, and Tewary 1992;
Canel, Carlsson, and Thomson 1995; Schi{\o}tz and Carlsson
1997)\nocite{ThZhCaTe92,CaCaTh95,ScCa97}.  In this method the atoms in
the outer zone are considered to interact through linear spring
forces, and the force matrix is inverted giving a lattice Green's
function.  The Green's function thus describes the displacement of the
lattice at one point resulting from a force acting on the lattice at
another point.  The degrees of freedom associated with the atoms in
the outer zone (typically a few million atoms) can then be
eliminated, and only the degrees of freedom associated with the atoms
in the inner zone need to be retained (typically less than 1000),
together with the elements of the Green's function matrix connecting
atoms in the inner zone.  The method has been used to study
dislocation emission from sharp and blunt cracks (Zhou, Carlsson, and
Thomson 1993,1994; Schi{\o}tz, Canel, and Carlsson
1997)\nocite{ZhCaTh93,ZhCaTh94,ScCaCa97}.  A similar
technique was used by Rao, Hernandez, Simmons, Parthasarathy and
Woodward (1998)\nocite{RaHeSiPaWo98} to study dislocation core
structures.

The above-mentioned techniques have the disadvantage that the system
must \emph{a priori} be divided into two ``zones'', and dislocations
and cracks are confined to the inner zone.  An adaptive technique has
been developed (Tadmor, Ortiz and Phillips 1996; Shenoy, Miller,
Tadmor, Rodney, Phillips and Ortiz, to be
published)\nocite{TaOrPh96,ShMiTaRoPhOrPREP}.  In this method the
system is described by a finite element method, but instead of an
ordinary constitutive relation, the energy is calculated from a tiny
atomistic simulation.  Where necessary, the finite element mesh is
automatically refined, eventually continuing until it is so fine
that each element only contains one atom.  In this limit the
simulation converges to an ordinary atomistic simulation.  The method
is adaptive, so the mesh is automatically refined in areas where a
detailed description is required, and coarsened where less detail is
required.  The method has been used to study the interactions of
dislocations and cracks with grain boundaries (Shenoy, Miller, Tadmor,
Phillips and Ortiz 1998; Shenoy \emph{et al.}, to be
published)\nocite{ShMiTaPhOr98,ShMiTaRoPhOrPREP}.

\section{CONCLUSIONS}

We have demonstrated how atomic-scale simulations can be used to
obtain information about the deformation mechanisms in nanocrystalline
metals.  We find that at these grain sizes the dominating deformation
mechanism is sliding in the grain boundaries through a large number of
small events in the grain boundary.  Each event only involves a few
atoms.  The result of this large number of small events is a flow in
the grain boundaries, permitting the grains to slide past each other
with only a minor amount of deformation inside the grains.  The
deformation mechanism results in a reverse Hall-Petch effect, where
the material becomes softer when the grain size is reduced. This is
caused by an increase of fraction of the atoms that are in the grain
boundaries as the grain size is reduced.  We have not yet been able
to observe the cross-over between the range of grain sizes where this
deformation mechanism dominates, and grain sizes where a conventional
deformation mechanism based on  dislocation motion dominates.  The
hardness and yield stress of nanocrystalline metals is expected to
reach its maximum in this cross-over region.

In order to simulate larger systems than the ones considered here, the
atomic-scale approach must be abandoned.  We have provided an overview
of simulation techniques at coarser scales, and have indicated how
they can be combined with atomic-scale simulations to provide models
of the mechanics of metals at multiple length scales.  These
techniques have not yet become mainstream simulation tools, but it is
likely that many of them will gain more widespread usage in the near
future.

\section{ACKNOWLEDGMENTS}

Center for Atomic-scale Materials Physics is sponsored by the Danish
National Research Council.  The present work was in part financed by
The Danish Technical Research Council through Grant No.~9601119.
Parallel computer time was financed by the Danish Research Councils
through Grant No.~9501775.

\section*{REFERENCES}
\setlength{\parskip}{0pt}
\renewcommand{\refname}{}
\renewcommand{\emph}{\uline}

\begin{thebibliography}{70}
\expandafter\ifx\csname natexlab\endcsname\relax\def\natexlab#1{#1}\fi

\bibitem[Abraham(1997)]{Ab97}
Abraham, F.~F. (1997).
\newblock On the transition from brittle to plastic failure in breaking a
  nanocrystal under tension ({NUT}).
\newblock Europhys. Lett. \emph{38}, 103--106.

\bibitem[Abraham et~al.(1997)Abraham, Schneider, Land, Lifka, Skovira, Gerner,
  and Rosenkrantz]{AbScLaLiSkGeRo97}
Abraham, F.~F., Schneider, D., Land, B., Lifka, D., Skovira, J., Gerner, J.,
  and Rosenkrantz, M. (1997).
\newblock Instability dynamics in three-dimensional fracture: an atomistic
  simulation.
\newblock J. Mech. Phys. Solids \emph{45}, 1461--1471.

\bibitem[Barts and Carlsson(1995)]{BaCa95}
Barts, D.~B. and Carlsson, A.~E. (1995).
\newblock Order-{$N$} method for force calculation in many-dislocation systems.
\newblock Phys. Rev. E \emph{52}, 3195.

\bibitem[Barts and Carlsson(1997)]{BaCa97}
Barts, D.~B. and Carlsson, A.~E. (1997).
\newblock Simulation and theory of polygonization in single glide.
\newblock Phil. Mag. A \emph{75}, 541--562.

\bibitem[Bulatov et~al.(1998)Bulatov, Abraham, Kubin, Devincre, and
  Yip]{BuAbKuDeYi98}
Bulatov, V., Abraham, F.~F., Kubin, L., Devincre, B., and Yip, S. (1998).
\newblock Connecting atomistic and mesoscale simulations of crystal plasticity.
\newblock Nature \emph{391}, 669--672.

\bibitem[Canel et~al.(1995)Canel, Carlsson, and Thomson]{CaCaTh95}
Canel, L.~M., Carlsson, A.~E., and Thomson, R. (1995).
\newblock Efficient effective-energy method for lattice-{G}reen's-function
  simulations of fracture.
\newblock Phys. Rev. B \emph{52}, 158--167.

\bibitem[Carlsson and Thomson(1998)]{CaTh98}
Carlsson, A.~E. and Thomson, R. (1998).
\newblock Fracture toughness of materials: From atomistics to continuum theory.
\newblock Solid State Physics \emph{51}, 233--280.

\bibitem[Chen(1995)]{Ch95}
Chen, D. (1995).
\newblock Structural modeling of nanocrystalline materials.
\newblock Comput. Mater. Sci. \emph{3}, 327--333.

\bibitem[Chokshi et~al.(1989)Chokshi, Rosen, Karch, and Gleiter]{ChRoKaGl89}
Chokshi, A.~H., Rosen, A., Karch, J., and Gleiter, H. (1989).
\newblock On the validity of the {H}all-{P}etch relationship in nanocrystalline
  materials.
\newblock Scripta Metall. \emph{23}, 1679--1684.

\bibitem[Clarke and J\'onsson(1993)]{ClJo93}
Clarke, A.~S. and J\'onsson, H. (1993).
\newblock Structural changes accompanying densification of random-sphere
  packings.
\newblock Phys. Rev. E \emph{47}, 3975--3984.

\bibitem[D'Agostino and Van~Swygenhoven(1996)]{AgSw96}
D'Agostino, G. and Van~Swygenhoven, H. (1996).
\newblock Stuctural and mechanical properties of a simulated nickel nanophase.
\newblock Mat. Res. Soc. Symp. Proc. \emph{400}, 293--298.

\bibitem[Daw and Baskes(1984)]{DaBa84}
Daw, M.~S. and Baskes, M.~I. (1984).
\newblock Embedded-atom method: Derivation and application to impurities,
  surfaces, and other defects in metals.
\newblock Phys. Rev. B \emph{29}, 6443--6453.

\bibitem[Devincre and Kubin(1994)]{DeKu94}
Devincre, B. and Kubin, L.~P. (1994).
\newblock Three dimensional simulations of plasticity.
\newblock In: Strength of Materials, edited by H.~Oikawa, K.~Maruyama,
  S.~Takeuchi, and M.~Yamaguchi (The Japan Institute of Metals). pp. 179--182.

\bibitem[Devincre and Kubin(1997)]{DeKu97}
Devincre, B. and Kubin, L.~P. (1997).
\newblock Mesoscopic simulations of dislocations and plasticity.
\newblock Mater. Sci. Eng. A \emph{234-236}, 8--14.

\bibitem[Finnis and Sinclair(1984)]{FiSi84}
Finnis, M.~W. and Sinclair, J.~E. (1984).
\newblock A simple empirical {$N$}-body potential for transition metals.
\newblock Phil. Mag. A \emph{50}, 45--55.

\bibitem[Finnis and Sinclair(1986)]{FiSi86}
Finnis, M.~W. and Sinclair, J.~E. (1986).
\newblock Erratum.
\newblock Phil. Mag. A \emph{53}, 161.

\bibitem[Furukawa et~al.(1996)Furukawa, Horita, Nemoto, Valiev, and
  Langdon]{FuHoNeVaLa96}
Furukawa, M., Horita, Z., Nemoto, M., Valiev, R.~Z., and Langdon, T.~G. (1996).
\newblock Microhardness measurements and the {H}all-{P}etch relationship in an
  {Al--Mg} alloy with submicrometer grain size.
\newblock Acta Mater. \emph{44}, 4619--4629.

\bibitem[Gulluoglu et~al.(1989)Gulluoglu, Srolovitz, LeSar, and
  Lomdahl]{GuSrLeLo89}
Gulluoglu, A.~N., Srolovitz, D.~J., LeSar, R., and Lomdahl, P.~S. (1989).
\newblock Dislocation distributions in two dimensions.
\newblock Scr. Met. \emph{23}, 1347--1352.

\bibitem[Gumbsch(1995)]{Gu95}
Gumbsch, P. (1995).
\newblock An atomistic study of brittle fracture: Toward explicit failure
  criteria from atomistic modeling.
\newblock J. Mater. Res. \emph{10}, 2897--2907.

\bibitem[Hall(1951)]{Ha51}
Hall, E.~O. (1951).
\newblock The deformation and ageing of mild steel: {III} {D}iscussion of
  results.
\newblock Proc. Phys. Soc. London \emph{B64}, 747--753.

\bibitem[Holt(1970)]{Ho70}
Holt, D.~L. (1970).
\newblock Dislocation cell formation in metals.
\newblock J. Appl. Phys. \emph{41}, 3197--3201.

\bibitem[Horita et~al.(1996)Horita, Smith, Furukawa, Nemoto, Valiev, and
  Langdon]{HoSmFuNeVaLa96}
Horita, Z., Smith, D.~J., Furukawa, M., Nemoto, M., Valiev, R.~Z., and Langdon,
  T.~G. (1996).
\newblock An investigation of grain boundaries in submicrometer-grained
  {Al--Mg} solid solution alloys using high-resolution electron microscopy.
\newblock J. Mater. Res. \emph{11}, 1880--1890.

\bibitem[Jacobsen et~al.(1987)Jacobsen, N{\o}rskov, and Puska]{JaNoPu87}
Jacobsen, K.~W., N{\o}rskov, J.~K., and Puska, M.~J. (1987).
\newblock Interatomic interactions in the effective-medium theory.
\newblock Phys. Rev. B \emph{35}, 7423--7442.

\bibitem[Jacobsen et~al.(1996)Jacobsen, Stoltze, and N{\o}rskov]{JaStNo96}
Jacobsen, K.~W., Stoltze, P., and N{\o}rskov, J.~K. (1996).
\newblock A semi-empirical effective medium theory for metals and alloys.
\newblock Surf. Sci. \emph{366}, 394--402.

\bibitem[J\'onsson and Andersen(1988)]{JoAn88}
J\'onsson, H. and Andersen, H.~C. (1988).
\newblock Icosahedral ordering in the {L}ennard-{J}ones liquid and glass.
\newblock Phys. Rev. Lett. \emph{60}, 2295--2298.

\bibitem[Juul~Jensen(1997{\natexlab{a}})]{Ju97b}
Juul~Jensen, D. (1997{\natexlab{a}}).
\newblock Orientational Aspects of Growth during Recrystallization (Ris{\o},
  Roskilde).

\bibitem[Juul~Jensen(1997{\natexlab{b}})]{Ju97a}
Juul~Jensen, D. (1997{\natexlab{b}}).
\newblock Simulation of recrystallization microstructures and textures: Effects
  of preferential growth.
\newblock Metall. Mater. Trans. A \emph{28A}, 15--25.

\bibitem[Kohlhoff et~al.(1991)Kohlhoff, Gumbsch, and Fischmeister]{KoGuFi91}
Kohlhoff, S., Gumbsch, P., and Fischmeister, H.~F. (1991).
\newblock Crack propagation in b.c.c. crystals studied with a combined
  finite-element and atomistic model.
\newblock Phil. Mag. A \emph{64}, 851--878.

\bibitem[Kratochv\'\i{}l et~al.(1997)Kratochv\'\i{}l, Saxlov\`a, Devincre, and
  Kubin]{KrSaDeKu97}
Kratochv\'\i{}l, J., Saxlov\`a, M., Devincre, B., and Kubin, L.~P. (1997).
\newblock On the sweeping of dipolar loops by gliding dislocations.
\newblock Mater. Sci. Eng. A \emph{234-236}, 318--321.

\bibitem[Kubin et~al.(1992)Kubin, Canova, Condat, Devincre, Pontikis, and
  Br\'echet]{KuCaCoDePoBr92}
Kubin, L.~P., Canova, G., Condat, M., Devincre, B., Pontikis, V., and
  Br\'echet, Y. (1992).
\newblock Dislocation microstructures and plastic flow: a {3D} simulation.
\newblock Solid State Phenomena \emph{23-24}, 455--472.

\bibitem[Li et~al.(1994)Li, Sun, and Wang]{LiSuWa94}
Li, S., Sun, L., and Wang, Z. (1994).
\newblock A grain boundary model of {H}all-{P}etch relationship in
  nanocrystalline materials.
\newblock In: Strength of Materials, edited by H.~Oikawa, K.~Maruyama,
  S.~Takeuchi, and M.~Yamaguchi (The Japan Institute of Metals), pp. 873--876,
  pp. 873--876.

\bibitem[Lian et~al.(1993)Lian, Baudelet, and Nazarov]{LiBaNa93}
Lian, J., Baudelet, B., and Nazarov, A.~A. (1993).
\newblock Model for the prediction of the mechanical behaviour of
  nanocrystalline materials.
\newblock Mater. Sci. Eng. A \emph{172}, 23--29.

\bibitem[Lu and Sui(1993)]{LuSu93}
Lu, K. and Sui, M.~L. (1993).
\newblock An explanation to the abnormal {H}all-{P}etch relation in
  nanocrystalline materials.
\newblock Scripta Metall. Mater. \emph{28}, 1465--1470.

\bibitem[Mills et~al.(1995)Mills, J\'onsson, and Schenter]{MiJoSc95}
Mills, G., J\'onsson, H., and Schenter, G.~K. (1995).
\newblock Reversible work transition state theory: application to dissociative
  adsorption of hydrogen.
\newblock Surf. Sci. \emph{324}, 305--337.

\bibitem[Morris and Morris(1997)]{MoMo97}
Morris, D.~G. and Morris, M.~A. (1997).
\newblock Hardness, strength, ductility and toughness of nanocrystalline
  materials.
\newblock Mater. Sci. Forum \emph{235-238}, 861--872.

\bibitem[Nabarro(1987)]{Na87}
Nabarro, F. R.~N. (1987).
\newblock Theory of crystal dislocations (Dover, New York).

\bibitem[Nieh and Wadsworth(1991)]{NiWa91}
Nieh, T.~G. and Wadsworth, J. (1991).
\newblock {H}all-{P}etch relation in nanocrystalline solids.
\newblock Scripta Met. Mater. \emph{25}, 955--958.

\bibitem[Nieman et~al.(1990)Nieman, Weertman, and Siegel]{NiWeSi90}
Nieman, G.~W., Weertman, J.~R., and Siegel, R.~W. (1990).
\newblock Tensile strength and creep properties of nanocrystalline palladium.
\newblock Scripta Met. Mater. \emph{24}, 145--150.

\bibitem[Nieman et~al.(1991)Nieman, Weertman, and Siegel]{NiWeSi91}
Nieman, G.~W., Weertman, J.~R., and Siegel, R.~W. (1991).
\newblock Mechanical behavior of nanocrystalline {Cu} and {Pd}.
\newblock J. Mater. Res. \emph{6}, 1012--1027.

\bibitem[Pande et~al.(1993)Pande, Masumura, and Armstrong]{PaMaAr93}
Pande, C.~S., Masumura, R.~A., and Armstrong, R.~W. (1993).
\newblock Pile-up based {H}all-{P}etch relation for nanoscale materials.
\newblock Nanostruct. Mater. \emph{2}, 323--331.

\bibitem[Pedersen et~al.(1998)Pedersen, Carstensen, and Rasmussen]{PeCaRa98x}
Pedersen, O.~B., Carstensen, J.~V., and Rasmussen, T. (1998) These proceedings.

\bibitem[Petch(1953)]{Pe53}
Petch, N.~J. (1953).
\newblock The cleavage strength of polycrystals.
\newblock J. Iron Steel Inst. \emph{174}, 25.

\bibitem[Phillpot et~al.(1995{\natexlab{a}})Phillpot, Wolf, and
  Gleiter]{PhWoGl95}
Phillpot, S.~R., Wolf, D., and Gleiter, H. (1995{\natexlab{a}}).
\newblock Molecular-dynamics study of the synthesis and characterization of a
  fully dense, three-dimenstional nanocrystalline material.
\newblock J. Appl. Phys. \emph{78}, 847--860.

\bibitem[Phillpot et~al.(1995{\natexlab{b}})Phillpot, Wolf, and
  Gleiter]{PhWoGl95b}
Phillpot, S.~R., Wolf, D., and Gleiter, H. (1995{\natexlab{b}}).
\newblock A structural model for grain boundaries in nanocrystalline materials.
\newblock Scripta Met. Mater. \emph{33}, 1245--1251.

\bibitem[Press et~al.(1988)Press, Flannery, Teukolsky, and
  Vetterling]{PrFlTeVe88}
Press, W.~H., Flannery, B.~P., Teukolsky, S.~A., and Vetterling, W.~T. (1988).
\newblock Numerical recipes in {C} (Cambridge University Press, Cambridge).

\bibitem[Rao et~al.(1998)Rao, Hernandez, Simmons, Parthasarathy, and
  Woodward]{RaHeSiPaWo98}
Rao, S., Hernandez, C., Simmons, J.~P., Parthasarathy, T.~A., and Woodward, C.
  (1998).
\newblock Green's function boundary conditions in two-dimensional and
  three-dimensional atomistic simulations of dislocations.
\newblock Phil. Mag. A \emph{77}, 231--256.

\bibitem[Rasmussen et~al.(1997)Rasmussen, Jacobsen, Leffers, Pedersen,
  Srinivasan, and J\'onsson]{RaJaLePeSrJo97}
Rasmussen, T., Jacobsen, K.~W., Leffers, T., Pedersen, O.~B., Srinivasan,
  S.~G., and J\'onsson, H. (1997).
\newblock Atomistic determination of cross-slip pathway and energetics.
\newblock Phys. Rev. Lett. \emph{79}, 3676--3679.

\bibitem[Sanders et~al.(1997{\natexlab{a}})Sanders, Fougere, Thompson, Eastman,
  and Weertman]{SaFoThEaWe97}
Sanders, P.~G., Fougere, G.~E., Thompson, L.~J., Eastman, J.~A., and Weertman,
  J.~R. (1997{\natexlab{a}}).
\newblock Improvements in the synthesis and compaction of nanocrystalline
  materials.
\newblock Nanostruct. Mater. \emph{8}, 243--252.

\bibitem[Sanders et~al.(1997{\natexlab{b}})Sanders, Youngdahl, and
  Weertman]{SaYoWe97}
Sanders, P.~G., Youngdahl, C.~J., and Weertman, J.~R. (1997{\natexlab{b}}).
\newblock The strength of nanocrystalline metals with and without flaws.
\newblock Mater. Sci. Eng. A \emph{234-236}, 77--82.

\bibitem[Schi{\o}tz et~al.(1997)Schi{\o}tz, Canel, and Carlsson]{ScCaCa97}
Schi{\o}tz, J., Canel, L.~M., and Carlsson, A.~E. (1997).
\newblock Effects of crack tip geometry on dislocation emission and cleavage:
  {A} possible path to enhanced ductility.
\newblock Phys. Rev. B \emph{55}, 6211--6221.

\bibitem[Schi{\o}tz and Carlsson(1997)]{ScCa97}
Schi{\o}tz, J. and Carlsson, A.~E. (1997).
\newblock Calculation of elastic {G}reen's functions for lattices with
  cavities.
\newblock Phys. Rev. B \emph{56}, 2292--2294.
\newblock (BR).

\bibitem[Schi{\o}tz et~al.(1998)Schi{\o}tz, Di~Tolla, and Jacobsen]{ScDiJa98}
Schi{\o}tz, J., Di~Tolla, F.~D., and Jacobsen, K.~W. (1998).
\newblock Softening of nanocrystalline metals at very small grain sizes.
\newblock Nature \emph{391}, 561--563.

\bibitem[Schi{\o}tz et~al.(1995)Schi{\o}tz, Jacobsen, and Nielsen]{ScJaNi95}
Schi{\o}tz, J., Jacobsen, K.~W., and Nielsen, O.~H. (1995).
\newblock Kinematic generation of dislocations.
\newblock Phil. Mag. Lett. \emph{72}, 245--250.

\bibitem[Selitser and Morris(1994)]{SeMo94}
Selitser, S.~I. and Morris, J.~W. (1994).
\newblock Substructure formation during plastic deformation.
\newblock Acta Metall. Mater. \emph{42}, 3985--3991.

\bibitem[Shenoy et~al.(1998)Shenoy, Miller, Tadmor, Phillips, and
  Ortiz]{ShMiTaPhOr98}
Shenoy, V.~B., Miller, R., Tadmor, E.~B., Phillips, R., and Ortiz, M. (1998).
\newblock Quasicontinuum models of interfacial structure and deformation.
\newblock Phys. Rev. Lett. \emph{80}, 742--745.

\bibitem[Shenoy et~al.(to be published)Shenoy, Miller, Tadmor, Rodney,
  Phillips, and Ortiz]{ShMiTaRoPhOrPREP}
Shenoy, V.~B., Miller, R., Tadmor, E.~B., Rodney, D., Phillips, R., and Ortiz,
  M. (to be published).
\newblock An adaptive finite element approach to atomic-scale mechanics --- the
  quasicontinuum method (preprint cond-mat/9710027).

\bibitem[Siegel and Fougere(1994)]{SiFo94}
Siegel, R.~W. and Fougere, G.~E. (1994).
\newblock Mechanical properties of nanophase materials.
\newblock In: Nanophase Materials: Synthesis --- Properties --- Applications,
  edited by G.~C. Hadjipanayis and R.~W. Siegel (Kluwer, Dordrecht), vol. 260
  of \emph{NATO-ASI Series E: Applied Sciences}, pp. 233--261, pp. 233--261.

\bibitem[Stoltze(1997)]{St97}
Stoltze, P. (1997).
\newblock Simulation methods in atomic scale materials physics (Polyteknisk
  Forlag, Lyngby, Denmark).

\bibitem[Tadmor et~al.(1996)Tadmor, Ortiz, and Phillips]{TaOrPh96}
Tadmor, E.~B., Ortiz, M., and Phillips, R. (1996).
\newblock Quasicontinuum analysis of defects in solids.
\newblock Phil. Mag. A \emph{73}, 1529--1564.

\bibitem[Thomson et~al.(1992)Thomson, Zhou, Carlsson, and Tewary]{ThZhCaTe92}
Thomson, R., Zhou, S.~J., Carlsson, A.~E., and Tewary, V.~K. (1992).
\newblock Lattice imperfections studied by the use of lattice {G}reen's
  functions.
\newblock Phys. Rev. B \emph{46}, 10613--10622.

\bibitem[Valiev et~al.(1992)Valiev, Chmelik, Bordeaux, Kapelski, and
  Baudelet]{VaChBoKaBa92}
Valiev, R.~Z., Chmelik, F., Bordeaux, F., Kapelski, G., and Baudelet, B.
  (1992).
\newblock The {H}all-{P}etch relation in submicron-grained {Al}-1.5\%{Mg}
  alloy.
\newblock Scripta Metall. Mater. \emph{27}, 855--860.

\bibitem[Van~Swygenhoven and Caro(1997{\natexlab{a}})]{SwCa97b}
Van~Swygenhoven, H. and Caro, A. (1997{\natexlab{a}}).
\newblock Molecular dynamics computer simulation of nanophase {Ni}: structure
  and mechanical properties.
\newblock Nanostructured Materials \emph{9}, 669--672.

\bibitem[Van~Swygenhoven and Caro(1997{\natexlab{b}})]{SwCa97}
Van~Swygenhoven, H. and Caro, A. (1997{\natexlab{b}}).
\newblock Plastic behavior of nanophase {Ni}: a molecular dynamics computer
  simulation.
\newblock Appl. Phys. Lett. \emph{71}, 1652--1654.

\bibitem[Weertman(1993)]{We93}
Weertman, J.~R. (1993).
\newblock {H}all-{P}etch strengthening in nanocrystalline metals.
\newblock Mater. Sci. Eng. A \emph{166}, 161--167.

\bibitem[Zhou et~al.(1997)Zhou, Beazley, Lomdahl, and Holian]{ZhBeLoHo97}
Zhou, S.~J., Beazley, D.~M., Lomdahl, P.~S., and Holian, B.~L. (1997).
\newblock Large-scale molecular dynamics simulations of three-dimensional
  ductile failure.
\newblock Phys. Rev. Lett. \emph{78}, 479--482.

\bibitem[Zhou et~al.(1993)Zhou, Carlsson, and Thomson]{ZhCaTh93}
Zhou, S.~J., Carlsson, A.~E., and Thomson, R. (1993).
\newblock Dislocation nucleation and crack stability: Lattice
  {G}reen's-function treatment of cracks in a model hexagonal lattice.
\newblock Phys. Rev. B \emph{47}, 7710--7719.

\bibitem[Zhou et~al.(1994)Zhou, Carlsson, and Thomson]{ZhCaTh94}
Zhou, S.~J., Carlsson, A.~E., and Thomson, R. (1994).
\newblock Crack blunting effects on dislocation emission from cracks.
\newblock Phys. Rev. Lett. \emph{72}, 852--855.

\bibitem[Zhou et~al.(1998)Zhou, Preston, Lomdahl, and Beazley]{ZhPrLoBe98}
Zhou, S.~J., Preston, D.~L., Lomdahl, P.~S., and Beazley, D.~M. (1998).
\newblock Large-scale molecular dynamics simulation of dislocation intersection
  in copper.
\newblock Science \emph{279}, 1525--1527.

\bibitem[Zhu and Averback(1996)]{ZhAv96}
Zhu, H. and Averback, R.~S. (1996).
\newblock Sintering of nano-particle powders: simulations and experiments.
\newblock Materials and Manufacturing Processes \emph{11}, 905--923.

\bibitem[Zhu et~al.(1987)Zhu, Birringer, Herr, and Gleiter]{ZhBiHeGl87}
Zhu, X., Birringer, R., Herr, U., and Gleiter, H. (1987).
\newblock X-ray diffraction studies of nanometer-sized crystalline materials.
\newblock Phys. Rev. B \emph{35}, 9085--9090.

\end{thebibliography}

\end{document}